\documentclass[preprint2,numberedappendix]{emulateapj-rtx4}
\usepackage{graphicx,times,bm,url,color}
\graphicspath{{./fig/}{./png/}}

\newcommand{\EQ}{\begin{equation}}
\newcommand{\EN}{\end{equation}}
\newcommand{\EQA}{\begin{eqnarray}}
\newcommand{\ENA}{\end{eqnarray}}

\newcommand{\Sec}[1]{Section~\ref{#1}}

\newcommand{\Fig}[1]{Figure~\ref{#1}}

\newcommand{\meanrho}{\overline{\rho}}

{}
{}
{}

{}
{}
{}
{}
{}
{}
{}
{}
{}
{}
{}
{}
{}
{}
{}
{}
{}

{}
{}
{}
\newcommand{\meanA}{\overline{A}}
\newcommand{\meanB}{\overline{B}}

{}

{}
{}

\newcommand{\nullvector}{{\bf0}}

\newcommand{\uu}{\mbox{\boldmath $u$} {}}

\newcommand{\BB}{\mbox{\boldmath $B$} {}}

\newcommand{\AAA}{\mbox{\boldmath $A$} {}}

\newcommand{\nab}{\mbox{\boldmath $\nabla$} {}}

\def\la{\mathrel{\mathchoice {\vcenter{\offinterlineskip\halign{\hfil
$\displaystyle##$\hfil\cr<\cr\sim\cr}}}
{\vcenter{\offinterlineskip\halign{\hfil$\textstyle##$\hfil\cr<\cr\sim\cr}}}
{\vcenter{\offinterlineskip\halign{\hfil$\scriptstyle##$\hfil\cr<\cr\sim\cr}}}
{\vcenter{\offinterlineskip\halign{\hfil$\scriptscriptstyle##$\hfil\cr<\cr\sim\cr}}}}}

\def\Rm{\mbox{\rm Re}_M}

\def\betastar{\beta_{\star}}

\def\cs{c_{\rm s}}
\def\kf{k_{\it f}} 

\def\urms{u_{\rm rms}}

\def\etatz{\eta_{\rm t0}}

\def\tautd{\tau_{\rm td}}
\def\tauto{\tau_{\rm to}}

\def\meanBeq{\overline{B}_{\rm eq}}
\def\Beq{B_{\rm eq}}
\def\Beqz{B_{\rm eq0}}

\newcommand{\kG}{\,{\rm kG}}

\newcommand{\g}{\,{\rm g}}
\newcommand{\s}{\,{\rm s}}

\newcommand{\cm}{\,{\rm cm}}

\newcommand{\km}{\,{\rm km}}

\newcommand{\Mm}{\,{\rm Mm}}

\newcommand{\yr}{\,{\rm yr}}

\newcommand{\mins}{\,{\rm minutes}}
\newcommand{\days}{\,{\rm days}}
\newcommand{\hours}{\,{\rm hr}}

\newcommand{\yapj}[3]{ #1, {ApJ,} {#2}, #3}

\newcommand{\yapjl}[3]{ #1, {ApJ,} {#2}, #3}

\newcommand{\yan}[3]{ #1, {Astron.\ Nachr.,} {#2}, #3}

\newcommand{\yana}[3]{ #1, {A\&A,} {#2}, #3}

\newcommand{\ysovl}[3]{ #1, {Sov.\ Astron.\ Lett.,} {#2}, #3}
\newcommand{\yjetp}[3]{ #1, {Sov.\ Phys.\ JETP,} {#2}, #3}

\newcommand{\ymn}[3]{ #1, {MNRAS,} {#2}, #3}
\newcommand{\ynat}[3]{ #1, {Nature,} {#2}, #3}

\newcommand{\ysph}[3]{ #1, {Solar Phys.,} {#2}, #3}

\newcommand{\ypre}[3]{ #1, {Phys.\ Rev.\ E,} {#2}, #3}

\newcommand{\sapj}[2]{ #1, {ApJ}, submitted, arXiv:#2}
\newcommand{\sana}[2]{ #1, {A\&A}, submitted, arXiv:#2}

\received{2013 June 20}
\accepted{2013 September 19}
\begin{document}

\title{Self-assembly of shallow magnetic spots through strongly stratified turbulence}

\author{Axel Brandenburg$^{1,2}$, Nathan Kleeorin$^{3,1,4}$, and Igor Rogachevskii$^{3,1,4}$
}
\affil{
$^1$Nordita, Royal Institute of Technology and Stockholm University,
Roslagstullsbacken 23, 10691 Stockholm, Sweden\\
$^2$Department of Astronomy, AlbaNova University Center,
Stockholm University, 10691 Stockholm, Sweden\\
$^3$Department of Mechanical
Engineering, Ben-Gurion University of the Negev, POB 653,
Beer-Sheva 84105, Israel\\
$^4$Department of Radio Physics, N.~I.~Lobachevsky State University of
Nizhny Novgorod, Russia
}

\begin{abstract}
Recent studies have demonstrated that in fully developed turbulence,
the effective magnetic pressure of a large-scale field
(non-turbulent plus turbulent contributions) can become negative.
In the presence of strongly stratified turbulence,
this was shown to lead to a large-scale instability
that produces spontaneous magnetic flux concentrations.
Furthermore, using a horizontal magnetic field, elongated flux concentrations
with a strength of a few per cent of the equipartition value were found.
Here we show that a uniform {\em vertical} magnetic field leads to
circular magnetic spots of equipartition field strengths.
This could represent a minimalistic model of sunspot
formation and highlights the importance of two critical ingredients:
turbulence and strong stratification.
Radiation, ionization, and supergranulation may be important for realistic
simulations, but are not critical at the level of a minimalistic model
of magnetic spot formation.
\end{abstract}

\section{Introduction}

Over the last $30\yr$, there has been a growing consensus that sunspots are the
surface interceptions of long thin flux tubes that are anchored deep near
the bottom of the convection zone \citep{Par75,SW80,DSC93}.
By contrast, direct numerical simulations (DNS) of global convectively
driven dynamos produce a magnetic field that is distributed
throughout the convection zone \citep{BMBBT11}, either
with a maximum at the bottom of the convection
zone \citep{Racine} or at mid depths \citep{KMB12}.
Furthermore, while DNS have been able to
demonstrate the ascent of thin flux tubes within
a stratified layer \citep{Fan01}, convection simulations
such as those of \cite{GK11} have not produced evidence
that sufficiently strong tubes are a natural
result of a dynamo. On the contrary, once
simulations develop large-scale dynamo action,
they produce a more diffusive large-scale field
with a filling factor close to unity
\citep{KKB08}, which suggests that the large-scale field is more
densely packed and not in the form of thin tubes.
These types of arguments have led to the proposal
that the solar dynamo may be a distributed one
and that sunspots and active regions may be a
shallow phenomenon \citep{B05}.

The alternative scenario of a shallow origin of active regions
and sunspots faces difficulties too.
Simulations of \cite{Rem11} have shown that a realistic, sunspot-like
appearance of the magnetic field can be obtained when the field is
kept fixed at the bottom of the domain.
Related simulations have also been done for bipolar spots \citep{Cheung}.
Both studies emphasize the importance of radiative transfer.
While this is also true for the simulations of \cite{SN12},
they do demonstrate that keeping
the flux tubes fixed in space might not be needed
if the computational domain is big enough and new
horizontal field of 1\,kG is continuously supplied from the
bottom of their domain. They interpret their
findings in terms of magnetic flux being swept down and kept in
place at greater depth by the strong converging
flows associated with the supergranulation.
Yet another radiative magnetohydrodynamics simulation with realistic
physics is that of \cite{KKWM10}, who also find spontaneous flux
concentrations as a result of strongly converging flows, even though
their domain is more shallow and without supergranulation.
This work might also be related to that of \cite{TWBP98},
who showed that magneto-convection tends to segregate
into magnetized and unmagnetized regions.

The purpose of the present paper is to emphasize that the spontaneous
assembly of magnetic flux can be caused by purely hydromagnetic effects
without involving convection,
supergranulation, radiative transport, or even energy flux.
Our proposal is based on recent numerical evidence that the
so-called negative effective magnetic pressure instability (NEMPI),
which was originally discovered in analytical studies
by \cite{KRR89,KRR90}, does really work
\citep{BKKMR11,KBKR12,KBKMR12a,KBKMR12b}.
However, a serious problem with this approach was that in
the presence of an imposed horizontal magnetic field, the
strongest flux concentrations are typically just
some 10\% of the equipartition value. We now show
that this restriction is alleviated when there is
a small vertical net flux through the domain. In
that case the magnetic field arranges itself in
the form of a spot-like assembly that is fully
confined by the turbulent flow itself, without
the need for keeping it in place by artificial means.

\begin{figure*}\begin{center}
\includegraphics[width=\textwidth]{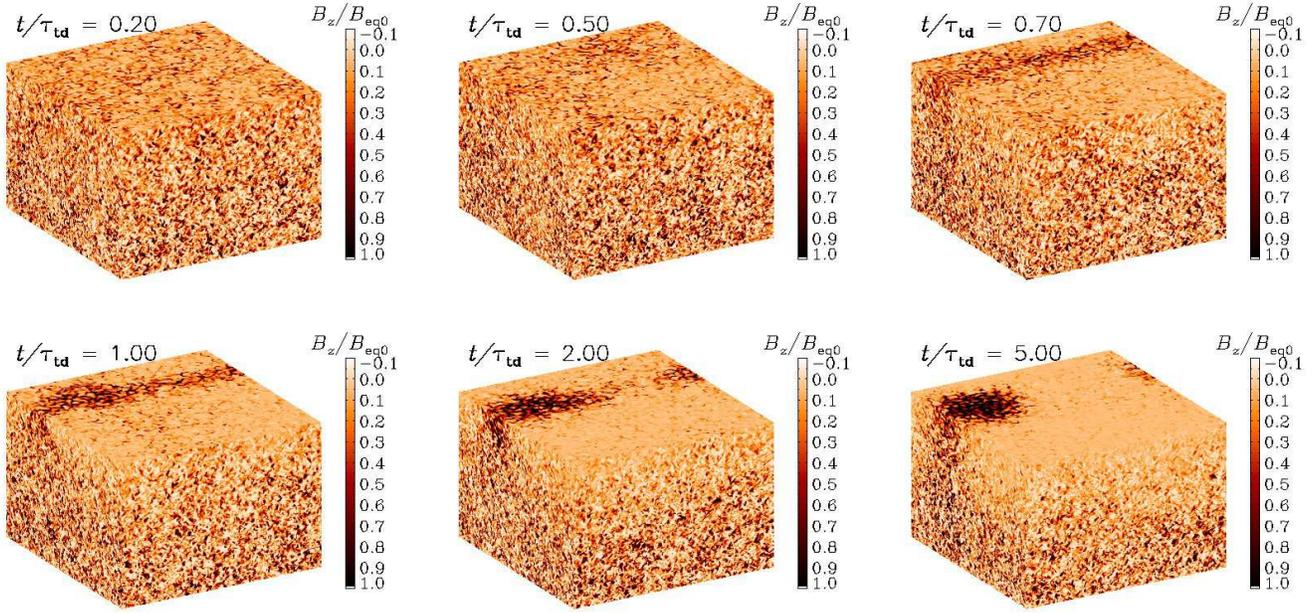}
\end{center}\caption[]{
Evolution from a uniform initial state toward a circular spot
for $B_{z0}/\Beqz=0.02$.
Here, $B_z/\Beqz$ is shown on the periphery of the domain.
Dark shades correspond to strong vertical fields.
Time is in units of $\tautd$.
An animation is available on {\tt http://youtu.be/Um\_7Hs\_1RzA}.
}\label{AB}\end{figure*}

The underlying mechanism of NEMPI is based on the suppression
of turbulent pressure and has been studied analytically \citep{KR94,RK07}
and numerically \citep{BKKR12,Losada1,Losada2,Jabbari},
using mean-field simulations (MFS) and DNS.
It can be understood as a negative contribution
of turbulence to the effective mean magnetic pressure
(the sum of non-turbulent and turbulent
contributions). At large Reynolds numbers
this turbulent contribution becomes large and NEMPI
can be excited.
The presence of strong density stratification
(small density scale height, $H_\rho$) is crucial,
because it leads to a negative magnetic buoyancy force.
(A local increase of the magnetic field causes a
decrease of the negative effective magnetic pressure,
which is compensated for by enhanced gas pressure,
leading to enhanced gas density,
so the gas is heavier than its surroundings and sinks.)
This results in a positive feedback loop:
downflow compresses the field,
the effective magnetic pressure becomes more negative,
gas pressure increases, so the density increases,
and the downflow accelerates;
see Equations~(4)--(9) of \cite{KBKMR12b} for
a phenomenological approach.
However, for magnetic fields close to equipartition,
the effective magnetic pressure becomes positive again,
so the instability saturates.
Significant scale separation between the forcing scale
and the size of the domain (about 15--30) is important, because
smaller turbulent eddies imply smaller turbulent diffusion;
see Figure~17 of \cite{BKKR12}.

\section{Details of the model}
\label{Differences}

Our goal is to present a minimalistic model
capable of producing a magnetic spot.
Within the framework of NEMPI, all that is needed is
turbulence, large enough scale separation, and strong stratification.
Our basic setup was described in \cite{BKKMR11,BKKR12}, where
non-helically driven turbulence was simulated in an isothermally
stratified domain.
In that case, $H_\rho$ is constant,
so the effects of strong stratification are distributed over all heights.
The forcing consists of random plane waves with constant amplitude,
so the rms velocity of the
turbulence, $\urms$, is independent of $z$.
As shown by \cite{KBKR12}, the theoretically expected maximum growth rate
of NEMPI, $\lambda_0\approx\betastar\urms/H_\rho$, is then the same for magnetic
fields at different heights, although the depth
where NEMPI is excited, increases with
increasing field strength.
Here, $\betastar$ is a non-dimensional parameter that was found to be
around 0.3 for magnetic Reynolds numbers in the range $1\la\Rm\la60$,
and 0.2 is for larger $\Rm$,
when small-scale dynamo action is possible \citep{BKKR12}.
Furthermore, in addition to isothermal stratification,
the equation of state is assumed isothermal, so the stabilizing
effects from Brunt-V\"ais\"al\"a oscillations are absent
\citep[see][where this has been relaxed in some of their MFS]{KaBKMR12}.
The scale separation ratio is taken to be 30, i.e., there are on average
30 turbulent cells across the domain; see \cite{Losada2} and \cite{Jab2},
where magnetic energy spectra are shown.

\begin{figure}\begin{center}
\includegraphics[width=\columnwidth]{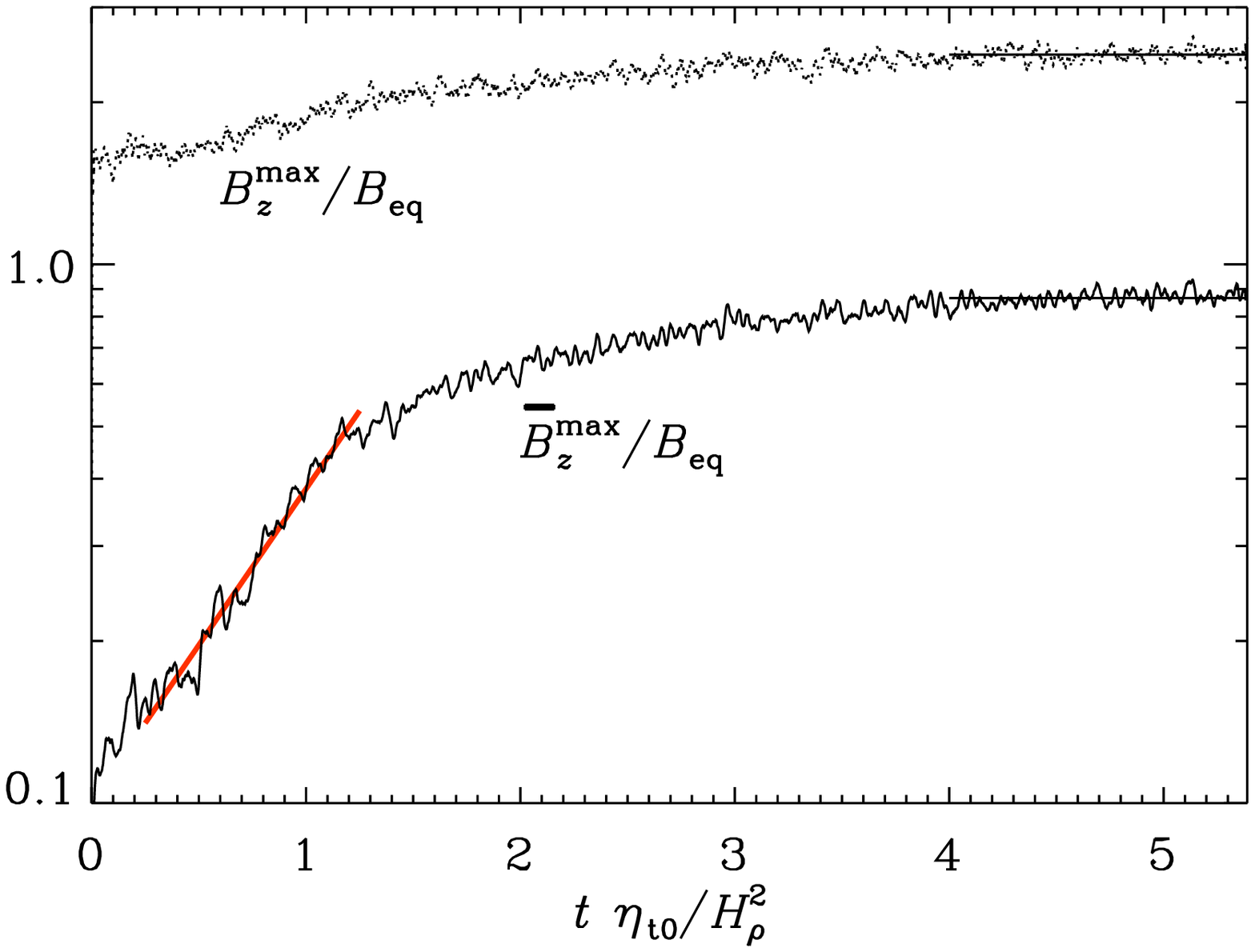}
\end{center}\caption[]{
Growth of $\meanB_z^{\max}/\Beq(z)$ (solid) and $B_z^{\max}/\Beq(z)$ (dotted)
at the top boundary.
The straight red line corresponds to a growth rate of $1.3\,\etatz/H_\rho^2$.
}\label{pBz_fit}\end{figure}

\begin{figure*}\begin{center}
\includegraphics[width=\textwidth]{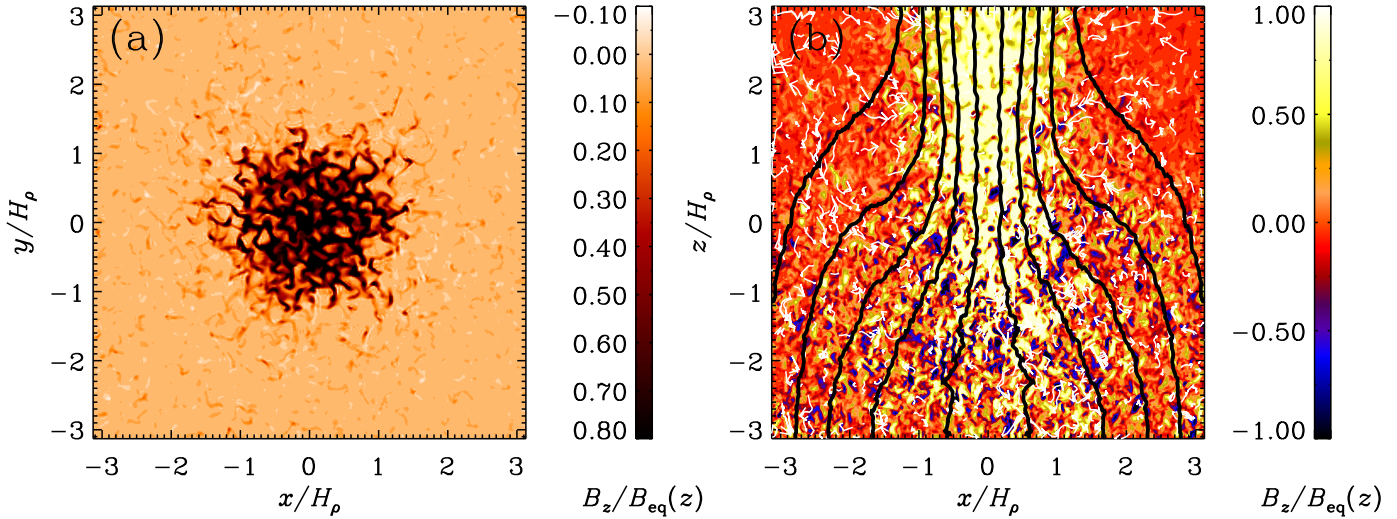}
\end{center}\caption[]{
Cuts of $B_z/\Beq(z)$ in the $xy$ plane at the top boundary
($z/H_\rho=\pi$) and the $xz$ plane through the middle of the spot at $y=0$.
In the $xz$ cut, we also show magnetic field lines and flow vectors
obtained by numerically averaging in azimuth around the spot axis.
}\label{pslices_V256k30VF_Bz002}\end{figure*}

The simulations are performed with the {\sc Pencil Code},%
\footnote{{\tt http://pencil-code.googlecode.com}}
which uses sixth-order explicit finite differences in space and a
third-order accurate time stepping method.
The magnetic field $\BB$ is
expressed in terms of the magnetic vector
potential $\AAA$ such that
$\BB=\BB_0+\nab\times\AAA$ is divergence-free and
$\BB_0=(0,0,B_0)$ is the imposed vertical field.
We use a numerical resolution of $256^3$ mesh
points in a Cartesian domain $(x,y,z)$ of size
$L^3$ such that $-L/2<x,y,z<L/2$. Our boundary
conditions are periodic in the horizontal
directions (so vertical magnetic flux is conserved), and stress free on the upper and
lower boundaries where the field is assumed to be
vertical, i.e., $B_x=B_y=0$.
Unless mentioned otherwise, the initial magnetic field is uniform
($\AAA=\nullvector$, so $\BB=\BB_0$) and our simulations are started from scratch.

Time is expressed in turbulent-diffusive times,
$\tautd=H_\rho^2/\etatz$, where
$\etatz=\urms/3\kf$ is the estimated turbulent
magnetic diffusivity and $\kf$ is the wavenumber
of the energy-carrying eddies.
Their turnover time is $\tauto=1/\urms\kf=\tau/\kf H_\rho$,
where $\tau=H_\rho/\urms$ is the natural time scale in a stratified layer.
Thus, $\tautd/\tau=3\,\kf H_\rho$ and $\tautd/\tauto=3\,(\kf H_\rho)^2$.
We use a setup that is similar to that of \cite{KBKMR12a},
where $H_\rho=1$ and $L=2\pi$,
so we have $L/H_\rho=2\pi\approx6$ scale heights across the domain.
The magnetic Reynolds number based on the wavenumber $\kf$
is $\Rm=\urms/\eta\kf\approx19$, with $\eta$ being the
microphysical magnetic diffusivity,
while that based on the scale $L$ is about 570.

The magnetic Prandtl number is 1/2 and the fluid Reynolds number is 38,
but simulations at resolutions of up to $1024^3$ and $\Rm=95$ give similar
results \citep{Jab2}.
The magnetic field is normalized by the local equipartition field
strength, $\Beq(z)=(\mu_0\meanrho)^{1/2}\urms$,
where $\mu_0$ is the vacuum permeability,
$\meanrho(z)$ is the horizontally averaged
density, while for $\urms$ we take the
root-mean-square based on a volume average,
because the turbulent velocity is driven such that it does not
show systematic height dependence.
(Since $B_0$ is small, the global $\urms$ does not change noticeably
during the simulation.)
We also define $\Beqz=\Beq(z=0)$ to specify the strength of the imposed
vertical magnetic field, as well as $\Beq(x)$ and $\meanBeq(x)$ to
characterize the local horizontal variation of $(\mu_0\rho\uu^2)^{1/2}$
through the magnetic spot.
Overbars denote Fourier filtering, as explained below.

\section{Results}

We have studied cases with different values of $B_0$.
We begin with $B_0/\Beqz=0.02$ and show in \Fig{AB}
the time evolution of the vertical magnetic field, $B_z$,
on the periphery of the domain.
Here, dark shades correspond to strong fields, so as to give an idea
how the temperature might look like if we relaxed the isothermal assumption.
Note in particular the gradual assembly of a magnetic spot from a uniform
turbulent background.
The color table is clipped for field strengths above the equipartition
value, while the field peaks at twice this value.
The time required for the development a magnetic spot
is 2--5 turbulent diffusive times.

\begin{figure*}\begin{center}
\includegraphics[width=\textwidth]{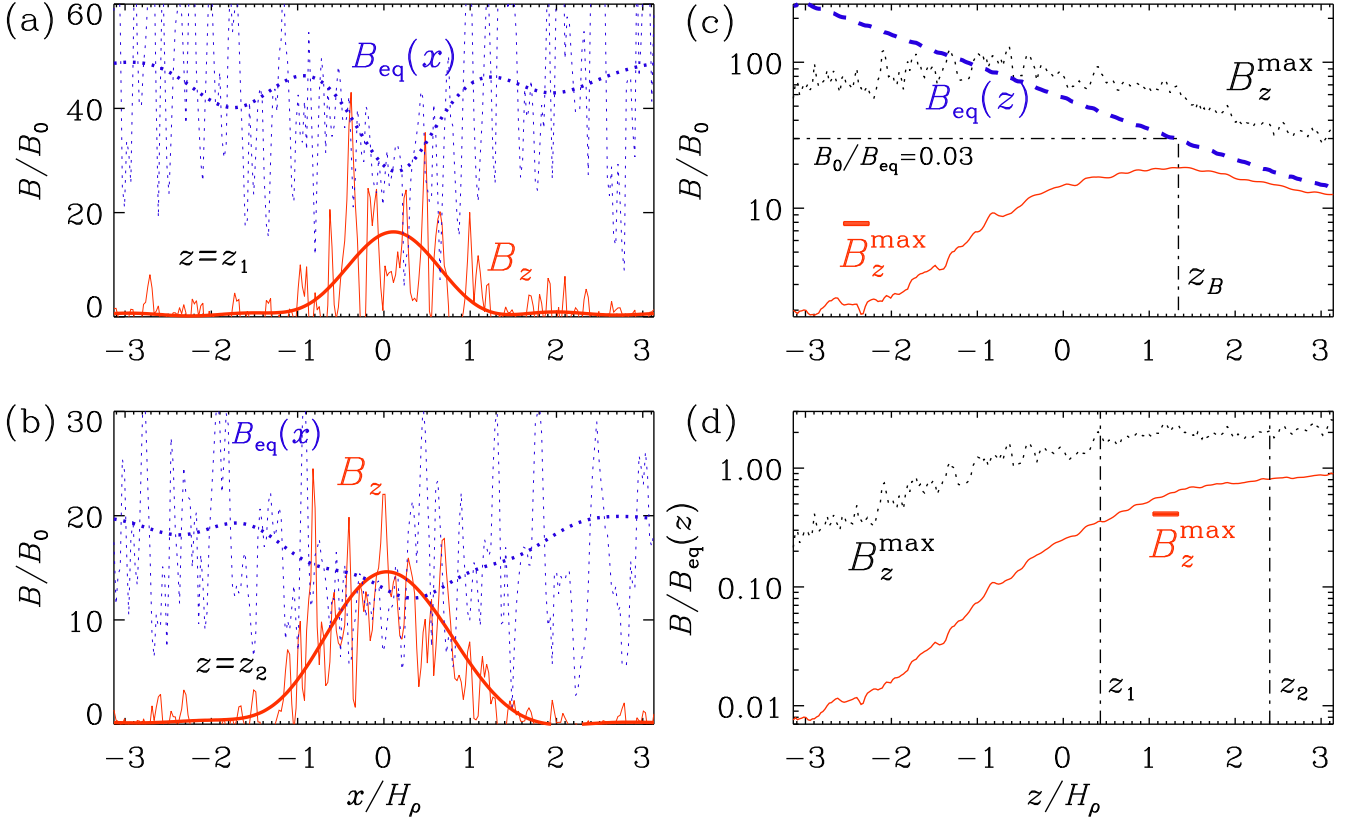}
\end{center}\caption[]{
Horizontal cross-tube profiles of $\Beq(x)$ and $B_z$,
where the smooth curves show respectively the spatial distribution of
$\meanBeq(x)$ and $\meanB_z$, normalized by $B_0$,
through $z=z_1\equiv0.4\,H_\rho$ (a) and $z=z_2\equiv2.4\,H_\rho$ (b),
and vertical profiles of $\meanB_z^{\max}$, $B_z^{\max}$, and $\Beq(x)$
normalized by $B_0$ (c) and by $\Beq(z)$ (d)
for the snapshot shown in \Fig{AB}.
Dash-dotted lines denote in (c)
the height $z_B$ where $B_0/\Beq(z_B)\approx0.03$,
and in (d) the positions $z_1$ and $z_2$.
}\label{pBprof_V256k30VF_Bz002_fresh}
\end{figure*}

The growth of the large-scale field is compatible with an exponential one
with a growth rate $\lambda\approx1.3\,\etatz/H_\rho^2$; see \Fig{pBz_fit},
where we show, at the top layer, the maximum field strength, $B_z^{\max}$, and the
maximum value of the large-scale field, $\meanB_z^{\max}$, obtained by
Fourier filtering to include only fields with horizontal wavenumbers below $\kf/6$.
Our value of $\lambda$ agrees with that of earlier studies of
magnetic flux concentrations in DNS
in the presence of a horizontal field and related MFS
of NEMPI \citep{KBKMR12a}.
However, our domain might not be large enough to include
the horizontal wavenumber $k_\perp$ of the fastest growing mode,
which has $k_\perp H_\rho\approx0.7$ \citep{Jab2}.
On the other hand, once the instability saturates, the magnetic field in the
direction of the imposed field gets more concentrated and
is then fully confined in the domain.
Between $t/\tautd=2$ and 5, it develops into
a nearly circular spot, as might be expected from
cylindrical symmetry arguments.

In \Fig{pslices_V256k30VF_Bz002} we show horizontal and vertical cuts
through the spot.
In the horizontal cut, again, strong fields correspond to dark shades.
The vertical cut is with a different color table where strong fields
now correspond to light shades.
It shows that the magnetic field (in units of the local equipartition
field strength) decreases with depth, but that fluctuations of
both signs (blue and yellow shades, respectively) become stronger.
We also consider the field averaged azimuthally about
the vertical axis of the tube.
Field lines correspond to contours of $\varpi\meanA_\theta(\varpi,z)$,
where $(\varpi,\theta,z)$ are cylindrical polar coordinates with
$\varpi$ being the cylindrical radius, $\theta$ the azimuthal angle,
and $z$ is identical to the Cartesian vertical coordinate.
This shows that the field in the tube fans out toward the bottom of the
domain and that the spot is only loosely anchored.

To analyze the magnetic spot quantitatively, we show in
\Fig{pBprof_V256k30VF_Bz002_fresh} horizontal and vertical cross-sections
for the snapshot shown in \Fig{pslices_V256k30VF_Bz002}.
It turns out that at some level $z=z_1$, the increase of $\meanB_z(x)$
at the position of the tube is matched by a corresponding decrease
in $\meanBeq(x)$; see \Fig{pBprof_V256k30VF_Bz002_fresh}(a).
We recall that $\Beq(x)$ was defined in \Sec{Differences} without averaging
so as to see its local suppression at the position of the spot.

At higher levels, only a small portion of the turbulent
kinetic energy is required to sustain the spot;
see \Fig{pBprof_V256k30VF_Bz002_fresh}(b) at $z=z_2$.
As a function of height, $\Beq(z)$ decreases monotonously.
Therefore, although the large-scale field representing the spot,
$\meanB_z^{\max}/B_0$, has a maximum at $z/H_\rho\approx1$,
as seen in \Fig{pBprof_V256k30VF_Bz002_fresh}(c), the field in
units of the equipartition field, $\meanB_z^{\max}/\Beq(z)$,
reaches a plateau in $1.5\la z/H_\rho\la3$;
see \Fig{pBprof_V256k30VF_Bz002_fresh}(d).
This is compatible with results by \cite{Losada3} that for a vertical field
the instability is strongest at a
height $z_B$ where $B_0/\Beq(z_B)\approx0.03$.
This is the case somewhere between $z_1$ and $z_2$, where
$\Beq(z)/B_0\approx30$ in \Fig{pBprof_V256k30VF_Bz002_fresh}(c).
Yet, in the nonlinear regime, near-equipartition field strengths are possible
in the upper part--both in DNS and the aforementioned MFS \citep{Jab2}.
Such sustained flux concentrations
might be assisted by slow inflows, as seen in the
upper part of \Fig{pslices_V256k30VF_Bz002}(b).

\begin{figure*}\begin{center}
\includegraphics[width=\textwidth]{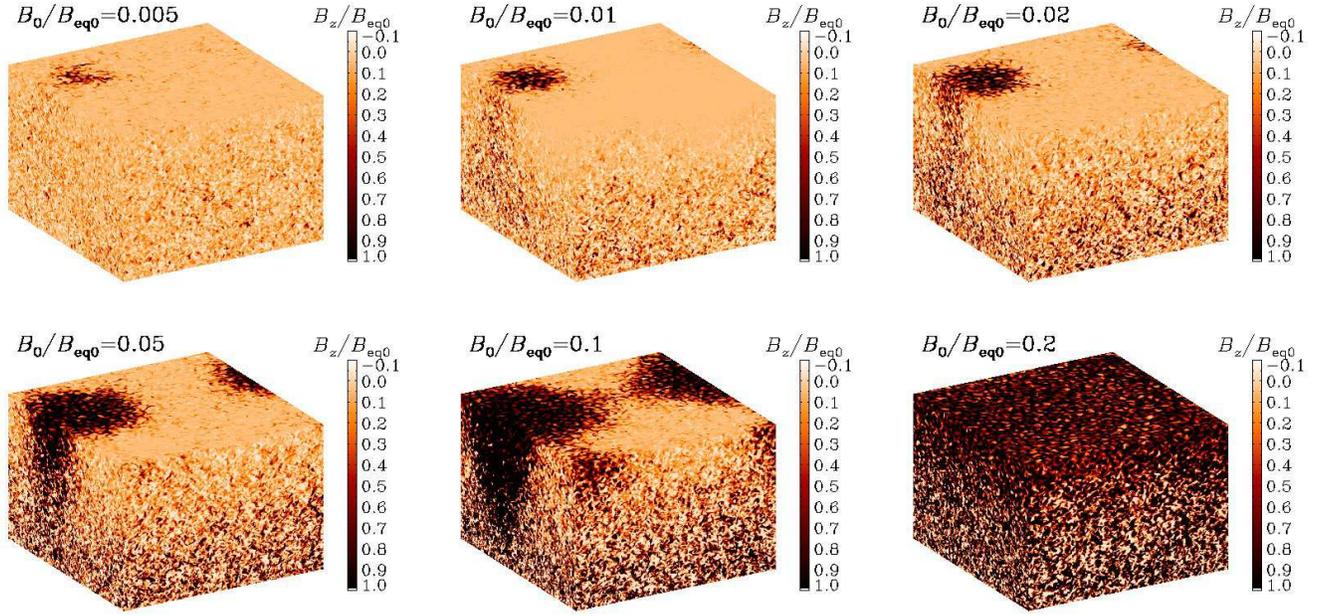}
\end{center}\caption[]{
Magnetic field structure for $B_0/\Beqz$ ranging from 0.005 to 0.2
showing the gradual transition from a small spot to a fully covered surface.
}\label{img_tlast}\end{figure*}

Finally, in \Fig{img_tlast} we compare simulations with different imposed
field strengths with $B_0/\Beqz$ from $5\times10^{-3}$ to 0.2.
The third panel corresponds to the last one of \Fig{AB},
which was also used as initial condition for the other runs.
The spot becomes smaller for weaker fields, while for stronger
fields the surface is eventually fully covered.
Note, however, that in all cases the large-scale field is approximately
of equipartition strength and roughly independent of the strength of
the imposed fields.
This is interesting in view of the fact that photospheric magnetic fields
of active stars are found to be of thermal equipartition strength
such that the filling factor grows as the star
becomes more active \citep{SL85}.

We emphasize that the magnetic field is not uniform across the spot, as is
assumed in a monolithic sunspot model, but it is more reminiscent of
the fibril sunspot model of \cite{Par79}.
In our case, there can even be regions where the field has the opposite sign.
This explains why in the last panel of \Fig{img_tlast} the field reaches
peak values above the equipartition value
over the entire horizontal plane---without
violating flux conservation, even though the large-scale field
is only 20\% of the equipartition value.
The value of $\urms/\cs\approx0.094$ is slightly less than
its original value of $\approx0.12$.

\section{Conclusions}

The present work has demonstrated two important aspects in the
production of magnetic flux concentrations:
the presence of a vertical magnetic field favors the formation of
circular structures and their field strengths can
exceed the local equipartition value.
The reason for such a strong effect in comparison with the case
with a horizontal imposed field is the apparent absence of
the so-called potato sack effect \citep[cf.][]{BKKMR11}.
We argue that this is a direct consequence of the negative
effective magnetic pressure, making such horizontal
magnetic structures heavier than their surroundings.
The potato sack effect is a nonlinear mechanism
responsible for a premature saturation of NEMPI with a horizontal field,
because it removes horizontal magnetic flux structures from
regions in which NEMPI is excited.
For a vertical magnetic field, the heavier fluid moves downward
along the field without affecting the flux tube, so that
NEMPI is not stabilized by the potato sack effect.
Instead, NEMPI saturates when $\meanB_z^{\max}/\Beq(z)=O(1)$;
see \Fig{pBprof_V256k30VF_Bz002_fresh}(d).

Application to the Sun is premature, but tentatively
we might estimate the time of spot formation
in solar values by using $\urms=1\km\s^{-1}$ and $H_\rho=300\km$,
so $\tau=H_\rho/\urms\approx5\mins$.
This, together with
$\kf H_\rho\approx2\pi\gamma/\alpha_{\rm mix}\approx6.5$
\citep{Losada2},
gives $\tautd\approx3\times6.5\times\tau\approx100\mins$,
and thus 5 turbulent diffusive times correspond to 8 hours
on a NEMPI length scale of $(2\pi/0.7)\times300\km\approx3\Mm$;
see \Sec{Differences}.
Here, $\gamma=5/3$ is the ratio of specific heats,
$\alpha_{\rm mix}=1.6$ is the mixing length parameter,
and $k_\perp H_\rho\approx0.7$ has been used \citep{Jab2}.
Conversely, at a depth where $H_\rho=3\Mm$, the length scale
would be $30\Mm$ and $5\tautd\approx80\hours\approx3\days$.
Furthermore, using $\rho=10^{-5}\g\cm^{-3}$, we have $\Beq\approx1\kG$,
so our model with $\meanB_z^{\max}/\Beq(z)\le1$ might
fall short of explaining the $3\kG$ field strengths observed in sunspots.

Our estimates are based on a minimalistic model of sunspot formation.
Nevertheless, these
new findings of flux concentrations with vertical fields
warrant further research in studying the origin of sunspots
and active regions.
Future developments include the addition of
(i) a radiating surface to move the top boundary condition
away from the upper boundary of the spot,
(ii) hydrogen ionization to allow for an extreme temperature
jump that might enhance the local growth of NEMPI,
(iii) dynamo-generated instead of imposed fields to allow
spots to come and go as the large-scale field evolves,
and finally (iv) convection instead of forced turbulence
to have a natural scale of turbulence with changes in its strength
in response to the magnetic field.

Studies involving radiation and ionization require the
solution of an energy equation, which might be important for
obtaining larger field strengths.
The simultaneous presence of NEMPI and dynamo instability has already
been studied in global MFS \citep{Jabbari} as well as in local
DNS in Cartesian geometry \citep{Losada2}.
The allowance for dynamo action is
particularly important from a morphological point of view.
It would give us a better idea about the appearance and
disappearance of spots, the possibility of bipolar regions \citep{WLBKR13},
and their inclination relative to the east--west direction,
which is expected due to the combined presence of poloidal
and toroidal fields in a dynamo.
Again, radiation might be important to allow
the field to develop more realistic inclinations about the vertical
and thereby also the formation of a penumbra.

\acknowledgments

We thanks the three anonymous reviewers for their detailed
comments and suggestions that have greatly improved the paper.
Computing resources provided by the Swedish National Allocations Committee
at the Center for Parallel Computers at the Royal Institute of Technology in
Stockholm and the High Performance Computing Center North in Ume{\aa}.
This work was supported in part by the European Research Council
under the AstroDyn Research Project No.\ 227952,
by the Swedish Research Council under the project grants
621-2011-5076 and 2012-5797, by EU COST Action MP0806,
by the European Research Council under the Atmospheric Research Project No.\
227915, and by a grant from the Government of the Russian Federation under
contract No. 11.G34.31.0048.


\end{document}